# Relaxation and emission of Bragg-mode and cavity-mode polaritons in a ZnO microcavity at room temperature

S. Faure[1], C. Brimont[1], T. Guillet[1], T. Bretagnon[1], B. Gil[1], F. Médard[2], D. Lagarde[2], P. Disseix[2], J. Leymarie[2], J. Zúñiga-Pérez[3], M. Leroux[3], E. Frayssinet[3], J.C. Moreno[3], F. Semond[3], S. Bouchoule[4]

1 : GES, UMR5650, Université Montpellier 2 – CNRS,
    Place E. Bataillon, F-34095 Montpellier, France.
2 : LASMEA, UMR 6602, UBP – CNRS,
    24 Avenue des Landais, F-63177 Aubière Cedex, France.
3 : CRHEA – CNRS,
    Rue Bernard Grégory, F-06560 Valbonne, France.
4 : LPN – CNRS,
    Route de Nozay, F-91460 Marcoussis, France.

**Abstract**

The strong coupling regime in a ZnO microcavity is investigated through room temperature photoluminescence and reflectivity experiments. The simultaneous strong coupling of excitons to the cavity mode and the first Bragg mode is demonstrated at room temperature. The polariton relaxation is followed as a function of the excitation density, showing a non thermal polariton distribution. A relaxation bottleneck is evidenced in the Bragg-mode polariton branch. It is partly broken under strong excitation density, so that the emission from this branch dominates the one from cavity-mode polaritons.

Semiconductor microcavities are attracting a large interest related to the bosonic properties and the large optical non-linearities of cavity-polaritons, which arise from the strong coupling of excitons and confined photons. Over the last decade, the developments of the polariton physics led to many remarkable demonstrations such as polariton condensation [1, 2], first evidenced for CdTe and GaAs microcavities at low temperatures. The strong coupling has then been achieved at room temperature with GaN microcavities [3], as well as polariton lasing [4, 5]. More recently, the interest for ZnO microcavities has increased due to the large oscillator strength and binding energy of ZnO excitons[6]: these characteristics provide the most robust polaritons for inorganic microcavities, stable at room temperature[7-9]. However, as the coupling between excitons and photons is stronger than in other materials, the polariton system becomes more complex. Our recent works [10, 11] have shown that the upper polariton branch is easily damped due to the excitonic absorption and its continuum, so that the corresponding resonance can only be observed in ZnO microcavities with a thin ($\lambda/4$) active layer. Moreover, the large Rabi energies recently reported in ZnO microcavities are evidenced by angle-resolved spectroscopy, with anti-crossings spreading over a wide angular range of typically 50° and a spectral range of 200 meV. These unusual magnitudes imply drastic requirements on the stop band of the distributed Bragg reflectors (DBRs) used for such cavities: for example, the broadly used AlN/(Al,Ga)N DBRs yield stop bands up to 400 meV wide[12]. As a consequence, even if the excitons are preferentially coupled to the cavity mode in such structures, their coupling to the first Bragg mode is far from negligible and may reach the strong coupling regime [13, 14].

In this letter we evidence the simultaneous strong coupling of excitons with a cavity-mode and a Bragg-mode in a ZnO hybrid microcavity at 300K. We study their relaxation and their emission. The angular distribution of the polaritons is out of equilibrium and typical of a so-called bottleneck regime for the Bragg-mode polariton, but this regime is partly broken as the exciton and polariton densities are increased. We discuss the interplay between both polariton branches and its consequences for the achievement of polariton lasing: under strong excitation, we show that the Bragg-mode polariton branch collects most of the polaritons after their relaxation, thus limiting the accumulation into cavity-mode polaritons.

The investigated ZnO microcavity consists in a $\lambda/4$ bulk ZnO active layer, embedded between an AlN/Al$_{0.3}$Ga$_{0.7}$N bottom DBR and a (Si,O)/(Si,N) top DBR [11, 15]. The nominal mirror reflectivities are 97% and 68% respectively. The growth of the 11-pair nitride DBR was realized by molecular beam epitaxy (MBE) on a (111)-oriented Si substrate [16], leading to a crack density of 10 to 100 mm$^{-1}$. The ZnO layer was grown by plasma-assisted MBE in a different reactor, and it is about 48 nm thick. Finally the top 5-pair dielectric DBR was obtained by plasma-enhanced chemical vapor deposition.

The sample was first studied by angle-resolved reflectivity at low and room-temperature, with an angular resolution of 1° and an energy resolution of 1 meV. Figure 1 presents the experimental spectra measured at T=300 K at a position on the sample where the cavity mode is nearly at zero-detuning with respect to the exciton transitions. The first Bragg mode is at negative detuning (-140 meV) at normal incidence. It becomes resonant with excitons for a detection angle of $\theta=40°$. For the present room temperature experiments, the A and B excitonic transitions, which are clearly resolved up to 77 K (not shown here), cannot be distinguished due to their homogeneous broadening [11]. In figure 2.a we compare the angle-dependent energies of reflectivity minima with the resonances calculated by the so-called coupled oscillator model. The Rabi splitting for the cavity-mode polariton around $\theta=0°$ is 40 meV at 300 K and reaches 60 meV at 77 K. It is slightly larger than the homogeneous broadening of the cavity mode (30 meV, corresponding to a quality factor Q≈100) and of the A and B excitons (25 meV at 300K). Around θ=40°, the signature of the strong coupling between excitons and the Bragg mode is less pronounced, with a Rabi splitting of about 30 meV. We attribute this smaller value to the reduced quality factor of the Bragg mode compared to the cavity mode. Larger Rabi splittings are expected for similar microcavities with thicker ZnO active layers (80-100 meV for a λ cavity with identical properties) but the demonstration of the strong coupling is then challenging due to the damping of the upper polariton branch by excitonic absorption [11]. The observed Rabi splitting is slightly smaller than the one measured for a similar cavity with a top metallic mirror [11], which is explained by the weaker photon confinement: the mode penetrates deeper in the DBR than in the metal thereby increasing the effective cavity length [17].

The coupled oscillator model treats cavity polaritons as a mixture of the different modes (cavity and Bragg modes, exciton resonances) considered as a basis of independent states. Figure 2.b presents the angle-dependent composition of the cavity-mode polariton branch. At θ=0°, the cavity-mode polariton at 3.30 eV presents a well balanced exciton-photon composition, whereas the Bragg-mode polariton at 3.18 eV has a small exciton content at θ=0° (Figure 2.c) due to the negative detuning.

The angle-resolved photoluminescence (PL) spectroscopy provides a different insight into the interplay between the cavity-mode and Bragg-mode polariton branches. Since the negative detuning of the Bragg mode is three times larger than the Rabi splitting, we may expect the polariton branches based on the cavity mode to be good candidates for the achievement of a non-linear emission because they are well isolated from other branches. However, we show below that the polaritons efficiently relax to the lower branch based on the Bragg mode, for which the emission is dominant under strong excitation density.

PL spectra (Figure 3) were obtained under quasi-resonant pulsed excitation at 3.49 eV, with an incidence angle of 40°, provided by the third harmonics of a Nd:YaG laser ($P_0$=0.1 mJ.cm$^{-2}$/pulse, pulse duration: 5ns, repetition rate 10Hz). By varying the detection angle (resolution of 3°), we measure the angular distribution of polariton emission and therefore the relative polariton occupancy of the different branches, which reflects the competition between relaxation and emission processes. Both cavity-mode and Bragg-mode polaritons are identified. The polaritons relax from the exciton reservoir through their interaction with phonons, at low excitation density, and also with excitons and other polaritons, when their density permits an efficient Coulomb interaction between particles [18].

At low excitation density ($P_0$ and 20 $P_0$), we observe that the cavity-mode polariton branch at 3.30 eV emits mainly around θ=0°, showing that the relaxation towards the minimum of this branch is efficient. On the contrary, Bragg-mode polaritons emit mainly at larger angles (typically 15° to 40° at the excitation density $P_0$), evidencing the so-called "bottleneck effect": their dispersion is too steep to allow an efficient phonon-mediated relaxation of the polaritons towards the bottom of the dispersion curve. The absence of a relaxation bottleneck for the cavity-mode polaritons at zero detuning has been recently predicted for similar ZnO microcavities [18]. Our observation of a bottleneck effect for the Bragg-mode polaritons at large negative detuning can be explained by their small exciton content, and the short photon lifetime: the polariton scattering rates are proportional to their excitonic component so that the relaxation of Bragg-mode polaritons is slower than the one of cavity-mode polaritons [19].

As the excitation power P is increased (Figures 3.c, 3.d), the angular distribution of the Bragg-mode polariton emission changes: the maximum of the emission shifts to $\theta=\pm20°$ at $P=20\,P_0$ and to $\theta=0°$ at $P=100\,P_0$. This suggests an acceleration of the polariton scattering processes, as the particle density increases: in spite of their strongly photonic character, Bragg-mode polaritons efficiently interact with the exciton reservoir and relax faster. However the deduced polariton occupancy (not shown here) does not reach a thermal regime with a given effective temperature, for both cavity-mode and Bragg-mode polaritons, due to the short photon lifetime [20]. For comparison, the partial suppression of the bottleneck occurs at exciton densities of the same order of magnitude as the ones previously measured for GaN microcavities [21], even though the excitation conditions are very different [22]. We should notice that the top DBR starts deteriorating at larger power densities, thus limiting the accessible range for the excitation density, as reported for GaN microcavities [5]. The absence of optical non-linearities in the investigated density range is related to the rather small quality factor of the present microcavity[18].

The relaxation to the Bragg-mode polariton branch is an important issue when producing a polariton accumulation in view of a non-linear emission. At large excitation density, the emission at $\theta=0°$ from Bragg-mode polaritons is much stronger than the one from cavity-polaritons (Figure 3.b). The present configuration of Bragg and cavity modes, i.e. a zero detuning for the cavity mode and a large negative detuning for the Bragg mode, is typical of most reported experimental results on ZnO microcavities [7-9]. We have shown here that the emission of such microcavities cannot be understood by assuming a relaxation from the exciton reservoir to a single lower polariton branch.

In conclusion, we have demonstrated the strong coupling regime at room temperature in a hybrid ZnO microcavity. We have observed the simultaneous strong coupling of excitons with the Bragg and cavity modes. The angle-resolved photoluminescence shows a competition between the emissions of both branches. At large excitation densities, the bottleneck effect on the Bragg-mode polariton branch is partly suppressed and its emission becomes much more intense than the one from cavity-polaritons. The microcavity emission should therefore be interpreted in a model including multiple polariton branches. This work also suggests that an efficient injection of polaritons into the sole cavity-mode polariton branch will require to improve the design of the DBRs, and especially to center the narrow stopband of the nitride DBR at lower energy.

We gratefully acknowledge Dr. G. Malpuech for fruitful discussions. The authors acknowledge financial support of ANR under "ZOOM" project Grant No. ANR-06-BLAN-0135.

# REFERENCES

[1]     R. M. Stevenson, V. N. Astratov, M. S. Skolnick, D. M. Whittaker, M. Emam-Ismail, A. I. Tartakovskii, P. G. Savvidis, J. J. Baumberg, and J. S. Roberts, Phys. Rev. Lett. **85**, 3680 (2000), http://link.aps.org/abstract/PRL/v85/p3680.

[2]     J. Kasprzak, M. Richard, S. Kundermann, A. Baas, P. Jeambrun, J. M. J. Keeling, F. M. Marchetti, M. H. Szymanska, R. André, J. L. Staehli, V. Savona, P. B. Littlewood, B. Deveaud and Le Si Dang, Nature **443**, 409 (2006), http://dx.doi.org/-10.1038/nature05131.

[3]     N. Antoine-Vincent, F. Natali, D. Byrne, A. Vasson, P. Disseix, J. Leymarie, M. Leroux, F. Semond, and J. Massies, Phys. Rev. B **68**, 153313 (2003), http://-dx.doi.org/ 10.1103/PhysRevB.68.153313.

[4]     S. Christopoulos, G. B. H. von Högersthal, A. J. D. Grundy, P. G. Lagoudakis, A. V. Kavokin, J. J. Baumberg, G. Christmann, R. Butte, E. Feltin, J.-F. Carlin, and N. Grandjean, Phys. Rev. Lett. **98**, 126405 (2007), http://link.aps.org/abstract/PRL/v98/-e126405.

[5]     G. Christmann, R. Butte, E. Feltin, J.-F. Carlin, and N. Grandjean, Appl. Phys. Lett. **93**, 051102 (2008), http://link.aip.org/link/?APL/93/051102/1.

[6]     M. Zamfirescu, A. Kavokin, B. Gil, G. Malpuech, and M. Kaliteevski, Phys. Rev. B **65**, 161205 (2002), http://link.aps.org/abstract/PRB/v65/e161205.

[7]     R. Shimada, J. Xie, V. Avrutin, U. Özgür, and H. Morkoç, Appl. Phys. Lett. **92**, 011127 (2008), http://dx.doi.org/ 10.1063/1.2830022.

[8]     M. Nakayama, S. Komura, T. Kawase, and D. Kim, J. Phys. Soc. Jpn. **77**, 093705 (2008), http://jpsj.ipap.jp/link?JPSJ/77/093705/.

[9]     R. Schmidt-Grund, B. Rheinländer, C. Czekalla, G. Benndorf, H. Hochmuth, M. Lorenz, and M. Grundmann, Appl. Phys. B **93**, 331 (2008), http://dx.doi.org/10.1007/-s00340-008-3160-x.

[10]    S. Faure, T. Guillet, P. Lefebvre, T. Bretagnon, and B. Gil, Phys. Rev. B **78**, 235323 (2008), http://link.aps.org/abstract/PRB/v78/e235323.

[11]    F. Médard, J. Zúñiga-Pérez, P. Disseix, M. Mihailovic, J. Leymarie, A. Vasson, F. Semond, E. Frayssinet, J. C. Moreno, M. Leroux, S. Faure, and T. Guillet, Phys. Rev. B **79**, 125302 (2009), http://link.aps.org/abstract/PRB/v79/e125302.

[12]    H. M. Ng, T. D. Moustakas, and S. N. G. Chu, Appl. Phys. Lett. **76**, 2818 (2000), http://link.aip.org/link/?APL/76/2818/1.

[13]    M. Richard, R. Romestain, R. André, and L. S. Dang, Appl. Phys. Lett. **86**, 071916 (2005), http://link.aip.org/link/?APL/86/071916/1.

[14]    G. Christmann, R. Butte, E. Feltin, A. Mouti, P. A. Stadelmann, A. Castiglia, J.F. Carlin, and N. Grandjean, Phys. Rev. B **77**, 085310 (2008), http://dx.doi.org/-10.1103/PhysRevB.77.085310.

[15]    The $\lambda/2$ cavity is in fact formed of the $\lambda/4$ ZnO layer and the first $\lambda/4$ AlGaN layer.

[16]    I. R. Sellers, F. Semond, M. Leroux, J. Massies, M. Zamfirescu, F. Stokker-Cheregi, M. Gurioli, A. Vinattieri, M. Colocci, A. Tahraoui, and A. A. Khalifa, Phys. Rev. B **74**, 193308 (2006), http://dx.doi.org/10.1103/PhysrevB.74.193308.

[17]    F. Réveret, P. Disseix, J. Leymarie, A. Vasson, F. Semond, M. Leroux, and J. Massies, Phys. Rev. B **77**, 195303 (2008), http://link.aps.org/abstract/PRB/v77/-e195303.

[18]    R. Johne, D. D. Solnyshkov, and G. Malpuech, Appl. Phys. Lett. **93**, 211105 (2008), http://link.aip.org/link/?APL/93/211105/1.

[19]  A. I. Tartakovskii, M. Emam-Ismail, R. M. Stevenson, M. S. Skolnick, V. N. Astratov, D. M. Whittaker, J. J. Baumberg, and J. S. Roberts, Phys. Rev. B **62**, R2283 (2000), http://link.aps.org/abstract/PRB/v62/pR2283.

[20]  J. Kasprzak, D. D. Solnyshkov, R. Andre, L. S. Dang, and G. Malpuech, Phys. Rev. Lett. **101**, 146404 (2008), http://link.aps.org/abstract/PRL/v101/e146404.

[21]  F. Stokker-Cheregi, A. Vinattieri, F. Semond, M. Leroux, I. R. Sellers, J. Massies, D. Solnyshkov, G. Malpuech, M. Colocci, and M. Gurioli, Appl. Phys. Lett. **92**, 042119 (2008), http://link.aip.org/link/?APL/92/042119/1.

[22]  Excitons in microcavity active layers have a short non-radiative lifetime $\tau_{non-rad}$ of the order of 10 ps. It is comparable to the pulse duration used in ref [21] to excite GaN microcavities ($T_{pulse}$=8 ps) and much shorter than the one of the present work. In the first case the exciton density is of the order of $A.E/E_0$, where A is the absorption of the microcavity, taken equal to (1-R), E~1 µJ.cm$^{-2}$ is the energy per pulse and $E_0$ is the laser energy. In the second case, the excitation can be considered as continuous during each 5 ns long pulse, so that the exciton density is of the order of $A.E.\tau_{non-rad}/(E_0.T_{pulse})$, with $E.\tau_{non-rad}/T_{pulse}$~ 5 µJ.cm$^{-2}$. The exciton density is therefore similar in both cases, of the order of a few $10^{17}$ cm$^{-3}$.

# FIGURES CAPTIONS

Figure 1 : (a) Angle-resolved reflectivity spectra of the ZnO microcavity (T=300K; TE polarization); (b) detailed view of the anti-crossing between the Bragg mode and the exciton resonance. The green, red and blue dashed lines correspond at 5° to the Bragg-mode polaritons, the cavity-mode polaritons, and the upper polariton branch respectively.

Figure 2 : (a) Comparison between the measured resonances (dots) and the eigenstates of the coupled oscillator model (full lines). The error bars represent the calculated homogeneous broadening of the polariton modes. (b,c) Angular dependence of the composition of the cavity-mode polariton (b) and the Bragg-mode polariton (c).

Figure 3 : (a,c,d) Angle-resolved PL spectra in TE polarization (false colors); (b) Corresponding PL spectra detected at $\theta=0°$.

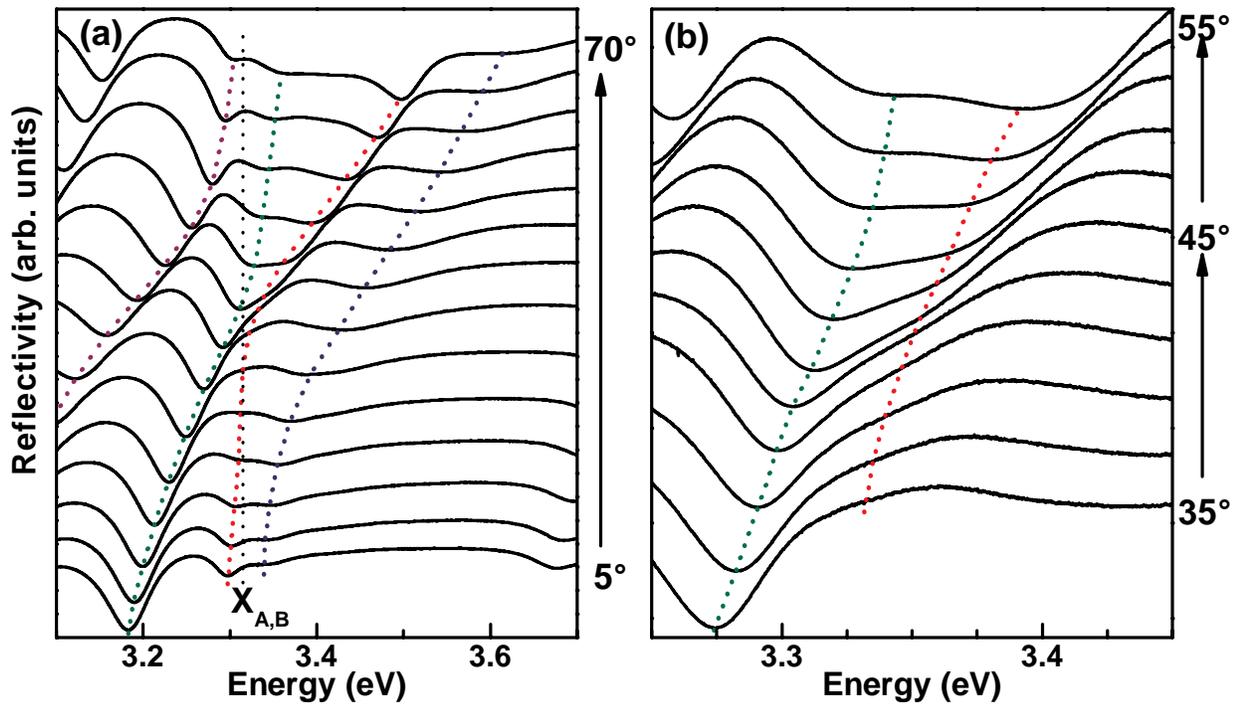

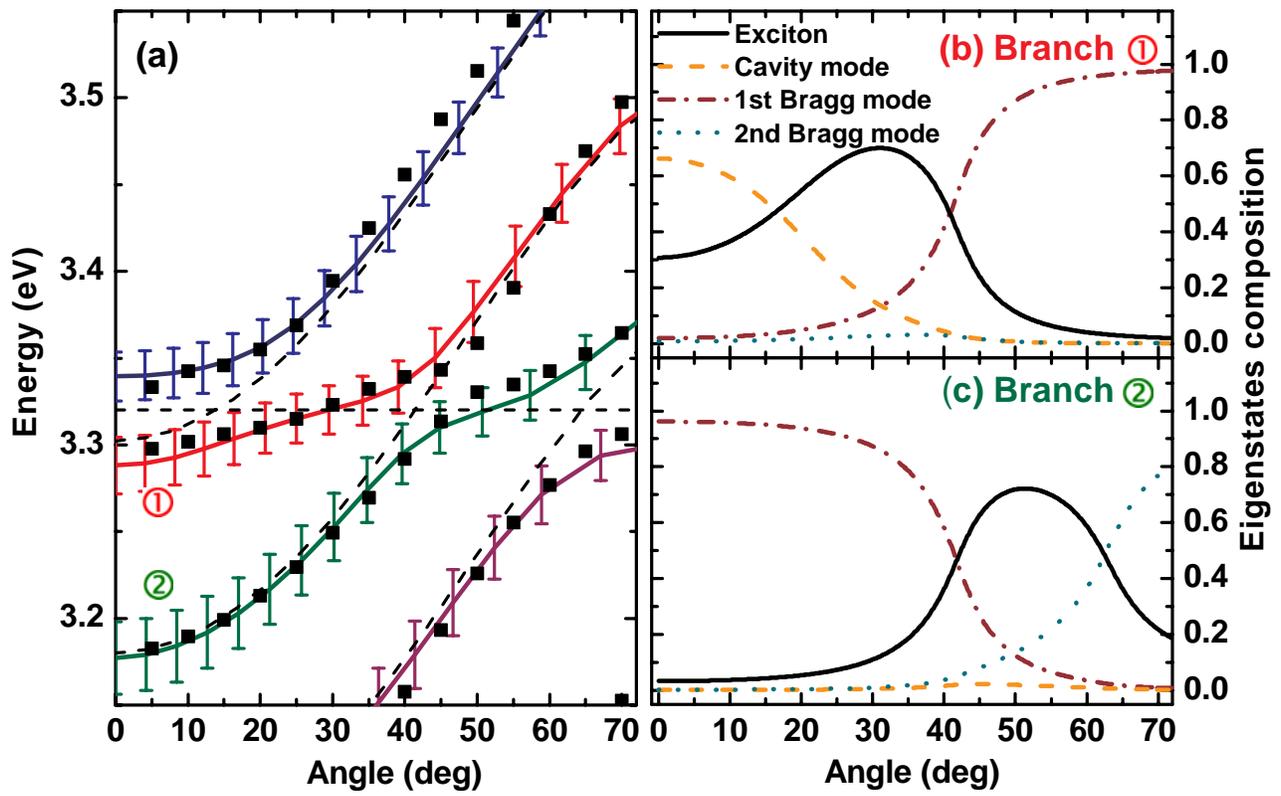

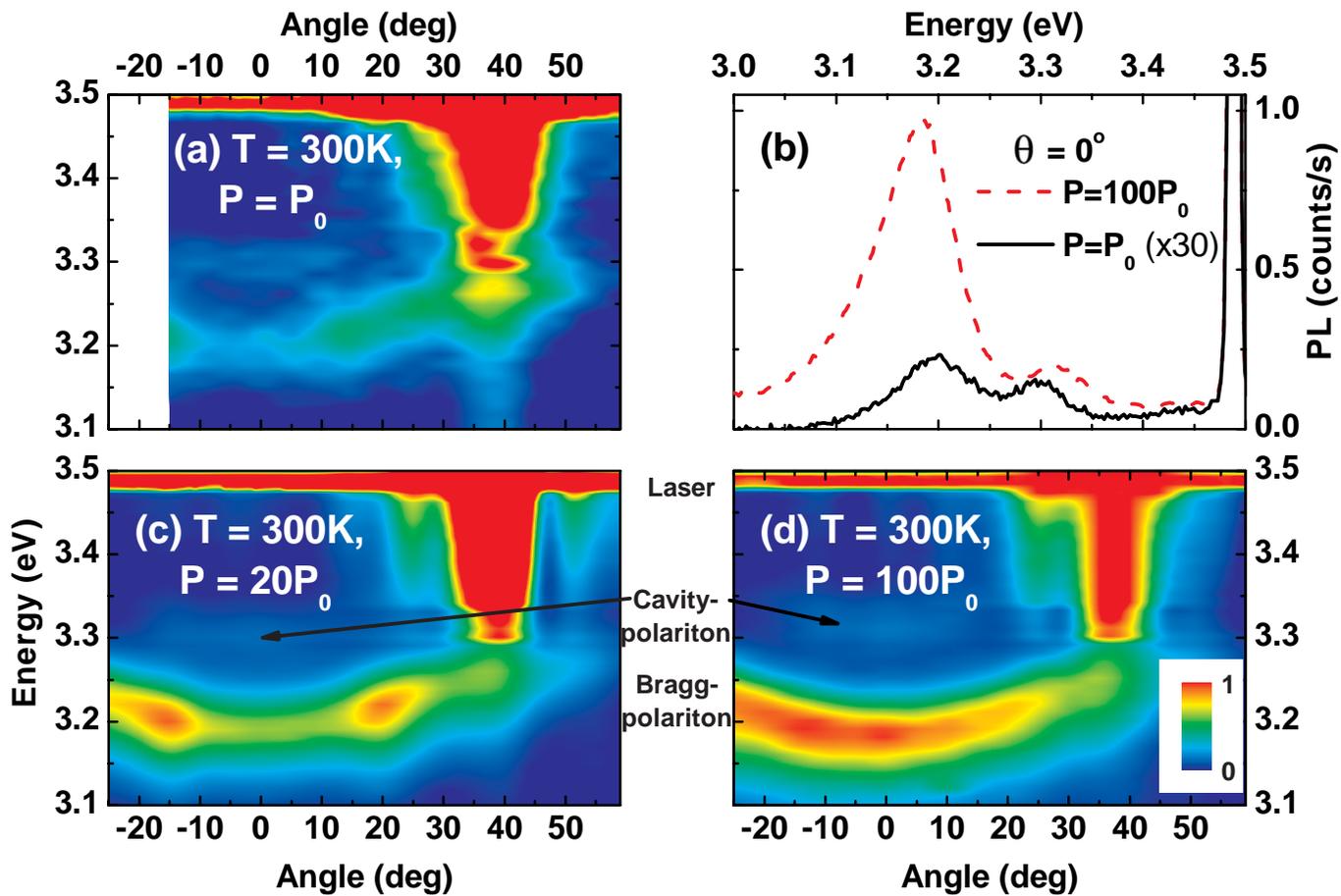